\documentclass[prl,
twocolumn,
preprintnumbers,noamssymb,aps]{revtex4}
\usepackage{epsf}
\usepackage{graphicx}
\usepackage{dcolumn}
\usepackage{amsmath}
\usepackage{color}
\def \be {\begin{equation}}
\def \ee {\end{equation}}
\def \bea {\begin{align}}
\def \eea {\end{align}}
\def \p {\partial}
\def \BEA {\begin{eqnarray}}
\def \EEA {\end{eqnarray}}
\def \BC {\begin{cases}}
\def \EC {\end{cases}}
\begin{document}
\title{Ratchet effect enhanced by plasmons}
\author{I.~V.~Rozhansky$^1$, V.~Yu.~Kachorovskii$^{1,2}$, and M.~S.~Shur$^2$}
\affiliation{$^1$A.F.~Ioffe Physical
Technical Institute, Russian Academy of Sciences, 194021
St.Petersburg, Russia \\
$^2$Center for  Integrated Electronics, Rensselaer Polytechnic Institute,  110, 8$^{th}$
Street, Troy, NY, 12180, USA }

\author{}
\affiliation{
Rensselaer Polytechnic Institute, Troy, USA
}


\date{\today}
\begin{abstract}
 {
Ratchet effect –-- a {\it dc} current  induced by the electromagnetic wave impinging on the
spatially modulated two-dimensional (2D) electron liquid  -- occurs when the wave
amplitude  is spatially modulated with the same wave vector as the 2D liquid but is
shifted in phase.  The analysis  within  the framework of the hydrodynamic model shows
that the ratchet current  is dramatically enhanced in the vicinity of the plasmonic
resonances and has nontrivial polarization dependence. In particular, for circular
polarization, the current component,  perpendicular to the modulation direction, changes
sign with the inversion of the radiation helicity. Remarkably, in the high-mobility
structures, this component  might be much larger than the  the current component  in the
modulation direction. We also discuss the non-resonant regime realized in dirty systems,
where the plasma resonances are suppressed, and demonstrate that the non-resonant
ratchet current is controlled by the Maxwell relaxation in the 2D liquid.
 }
\end{abstract}

\maketitle

Plasmonic oscillations in two-dimensional (2D) structures
have been
recently  a subject of a great interest in the context of the emerging field of plasma-wave electronics. The boost to this activity was given  about 20 years ago \cite{1}  by a theoretical prediction that a direct current ({\it dc}) in the channel of a field effect transistor (FET)
might become unstable with respect to  generation of plasma  oscillations.
    Such oscillations should lead to emission of radiation with the same frequency. It was also suggested  \cite{2} that
the  nonlinear properties of the  electron liquid in the FET channel
can be quite effectively used for rectifying  of the  plasma oscillation  induced by incoming electromagnetic wave.
    The  velocity of the plasma waves in the FET two-dimensional electron channel can be tuned by the gate voltage. Its typical value,
$\sim 10^8$ cm/s,  corresponds to the typical time scale of $10^{-12}$ s for the channel
length $\sim 1~\mu m$. Thus, a FET in the plasma waves regime is expected to
provide a tunable coupling to the electromagnetic radiation
in  the THz frequency range and can serve as a  THz emitter or
detector (for review   see Ref.~\onlinecite{15}).

There are, however, some difficulties in creating of such devices.  Since typical  FET dimensions are two or more orders of magnitude
smaller than THz wavelength, a single device weakly couples with the radiation.
 The coupling  dramatically increases
if there is a  {\it dc} current flowing in the FET channel
\cite{Veksler}. However, such current  leads to the increase of the device  noise.

Another possible way to increase coupling with the  radiation  is to use periodic structures (FET arrays, grating structures, and multi-gate structures)  instead of single FETs.  Such structures   attract  growing  interest   as   simple examples of plasmonic crystals \cite{Aizin1, Azin2, my1,  Azin3, Wang1}.    They are also very promising from point of view of possible applications and
     already demonstrated excellent performance as  THz detectors  \cite{allen1,allen2,allen4,28,allen6}, in a good agreement with numerical simulations \cite{32,30,31,popov}.   The first observations of   THz emission were also reported  \cite{29,new3}.

In this paper, we discuss  theoretically photo-response of a  FET array with a common channel and a large-area grating gate to the electromagnetic field. This structure represents a plasma crystal with a modulated gate potential.
 Non-zero response requires some asymmetry of the structure, which would
determine the direction of the produced {\it dc} current.   In a single FET, such asymmetry is induced  by asymmetrical boundary
conditions \cite{1}.
One of the possible   ways to induce  asymmetry in the plasma crystal  is related to     the so-called ratchet effect \cite{but1,but2,but5,but9,rachet1,rachet2, but13,but14,rachet4,but15}  (for review, see Refs.~\onlinecite{rachet1,rachet2,rachet4}).
Physically, the  rachet {\it dc} current   arises \cite{but13,but14, but15, but10,rachet4, but3, Golub, Popov13} as a result  of combined action   of a
static spatially-periodic in-plane potential (which can be created in a grating gate structures, see Fig.~1)
\begin{equation}\label{eq:V}
V(x) = V_0 \cos (qx)
\end{equation}
and   the electric field of incoming  radiation spatially modulated by a grating lattice (Fig.~1)  with the same $q$ \cite{comment0}:
\begin{equation}\label{eq:E}
\boldsymbol{ E}(t,x) =  \left[1+\hat h\cos (qx +\varphi)\right]\boldsymbol{ e} (t).
\end{equation}
 Here
 $\boldsymbol{ e} (t)=\big(e_x(t), e_y(t)\big)$ is in-plane oscillating vector with the components depending on   the polarization of the wave,  and $\hat h$ is diagonal $2\times 2$ matrix with the diagonal components $h_x$ and $h_y.$ These components describe the modulation depth of the radiation power in $x$ and $y$ directions, respectively.

The existence of non-zero average   $\big \langle \boldsymbol{E}  (\boldsymbol{E} \nabla  V) \big \rangle_{t,x} \propto \sin\varphi, $   implies that
  {\it dc} current $\boldsymbol{j}=(j_x,j_y) $     controlled by  the phase shift $\varphi$  between $V(x)$ and $\boldsymbol{E}(t,x)$ might appear   in the 2D liquid:
 $\boldsymbol{j} \propto \sin\varphi.$
 This phase shift serves
as the required asymmetry, so that the current reverses  its direction when $\varphi$ is shifted by $\pi.$
\begin{figure}[ht!]
   \leavevmode \epsfxsize=7.0cm
 \centering{\epsfbox{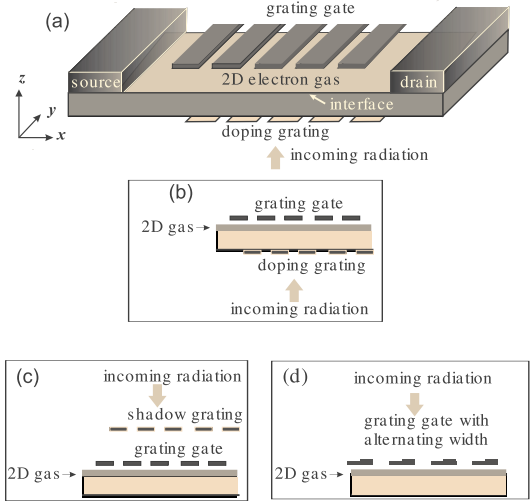}}
 \caption{Design of asymmetrical  grating gate structures.  Optical modulation
 can be achieved   by
  fabrication of doping grating  from the substrate  side  (a) [see     (b) for side view]
  or  shadow grating  from the gate side (c). Also one can use  grating gate
 that has alternating width and alternating transparency (d).}
\label{Fig1}
\end{figure}

The theory of the rachet effect  neglecting plasmonic effects  was developed in Refs.~\onlinecite{rachet1,rachet2,rachet4}.
It predicts  the Drude peak at zero frequency  of the radiation and otherwise a monotonic
smooth dependence of $\boldsymbol{j}$  on  $\omega$ in agreement with numerical simulations  \cite{Popov13}.

In this work, we demonstrate that excitation of plasmonic resonances  can dramatically increase the rectified {\it dc} current. We describe the plasmonic-enhanced  ratchet effect in the frame of the hydrodynamic model and obtain the
analytical expression for the {\it dc} current.  We demonstrate the  existence of the sharp plasmonic resonances in the dependence  $\boldsymbol{j}(\omega).$
The dependencies  $ j_x(\omega)$  and $ j_y(\omega)$ turn out to be different.
Remarkably,  in the high-mobility structures  the  component $ j_y(\omega)$ which is perpendicular to the modulation axis
might be much larger than $j_x.$
 The maximal  value of the ratio $j_y/j_x$  is achieved  in the vicinity of the  plasmonic resonances and is proportional to the quality factor of the structure.  Another intriguing property is the dependence of    $ j_y$ on the helicity of the polarization. For a single FET, the helicity-dependent response  was measured \cite{Ganichev}  and explained theoretically \cite{Romanov}  by assuming a  special type of the boundary conditions. The dependence of the $dc$ current   on the helicity in the grating-gate  periodic  structures   was also predicted in Refs.~\onlinecite{rachet4,Golub} within the approximation that ignores  plasmonic effects.     We will demonstrate that the helicity-dependent part of the response is also dramatically enhanced by the plasmonic effects.

We consider the   electron liquid in 2D channel in the external field \eqref{eq:E} of general polarization:
\be
{e_x} = {E_{0x}}\cos {\omega}t, ~~  {e_y} = {E_{0y}}\cos \left( {{\omega}t + \theta } \right)
 \ee
 The case $|E_{0x}|= |E_{0y}|$,  $\theta= \pm \pi/2$ corresponds to the
 circular polarization. For $E_{0y}=0$  the wave is linear polarized along  the  $x-$direction.
 In the absence of perturbations ($V=0,~\boldsymbol{E}=0$), the  2D electron concentration $N=N_0$
 is controlled by the gate-to-channel  voltage $U_g:$
\begin{equation}\label{eq:nx}
N_0 = \frac{{C U_g }}{e},
\end{equation}
where
$C=\varepsilon/4\pi d$
is the  gate-to-channel capacitance per unit area, $\varepsilon$  is the dielectric constant,
$d$ is the spacer distance, and  $e>0$  is the absolute value of the electron charge.
For smooth perturbations with $qd \ll 1$ equation \eqref{eq:nx} is also  valid  and  relates local concentration  in the channel $N=N(x,t)$ with the local gate-to-channel swing.
The total   electric field  in the channel is  given by  the sum of external field of radiation, static built-in field, and electrostatic field arising due to the density perturbation:  $ \boldsymbol{E}_{tot}=
\boldsymbol{E} -\nabla V +(e/C)\nabla N.$

 The quasiclassical dynamics of   electrons in the channel obeys kinetic equation:
\be \frac{\p f}{\p t}+ \mathbf {v}{\nabla} f+\left[\boldsymbol{a}- \frac{e^2}{mC}{\nabla}N \right]\frac{\p f}{\p \mathbf  v} = {\rm St} f,
\label{kin}
\ee
where
$ \boldsymbol {a}=-\frac{e}{m}\big(\boldsymbol{E}-  {\nabla} V\big),
$
 and ${\rm St} f$  is the collision integral including scattering off impurities and phonons as well as electron-electron scattering.   We  will  study electron liquid  within the hydrodynamic approximation  assuming the following hierarchy of the scattering times:
$\tau_{ee} \ll \tau \ll \tau_{ph},
$
 where $\tau_{ee},\tau $ and $ \tau_{ph}$  are the  electron-electron, impurities     and electron-phonon scattering times, respectively.   These  inequalities  allows one to search a solution  as a Fermi-Dirac function in the moving frame
  $f=
   1 \left /\left[e^{m(\mathbf v-\boldsymbol v)^2/2T - \mu/T}+1\right]. \right.
  $
    This function depends on the local hydrodynamic parameters: velocity $\boldsymbol v=\boldsymbol v(\mathbf r, t),$ chemical potential  $ \mu=\mu(\mathbf r, t),$   and temperature   $T=T(\mathbf r,t ).$  In what follows we set
      $\mu \gg T.$ This yields
      $N \approx \nu \mu,$
    where  $\nu=m/\pi\hbar^2$ is the density of states.   Having in mind that  the electron-electron collisions conserve the particle number, momentum and energy, we multiply Eq.~\eqref{kin}
 by  $1,$ $m\mathbf v$ and $m\mathbf v^2/2$ and integrate over momenta, thus  obtaining
  the system of coupled equations for hydrodynamic parameters:
  \BEA
&&
\label{cont00}
 \frac{{\partial N}}{{\partial t}} + \frac{\partial }{{\partial x}}\left( {N\boldsymbol{v}} \right) = 0,
\\&&\label{hydro00}
 \frac{{\partial \boldsymbol {v}}}{{\partial t}} + (\boldsymbol{ v}  {\nabla}) \boldsymbol{ v } + \frac{\boldsymbol{v}}{\tau } = \boldsymbol{a }  - \frac{{{e^2}}}{{mC}}  {\nabla} N - \frac{{\nabla} W}{mN}  ,
\\&&
{\cal C}\left [ \frac{\p T}{\p t} + {\rm div} ({T\boldsymbol{v}})\right]= N \left( \frac{T_0-T}{\tau_{ph}}+\frac{mv^2}{\tau} \right), \label{T}
\EEA
where
    $W= \int d\epsilon {\nu \epsilon}[e^{(\epsilon-\mu)/T} +1]^{-1}
  \approx {N^2}/{2\nu} + {\nu T^2\pi^2}/{6}$
    is the system energy per unit area in the moving frame, $T_0$ is the lattice temperature    and
 ${\cal C}=\nu T \pi^2/3$ is the heat capacity of  the 2D degenerate electrons. In above, we implicitly assumed that $\tau$  is energy independent, which is the case for  the impurity potential modeled by short-range disorder.

Equation \eqref{T} is coupled to Eqs.~\eqref{hydro00} and \eqref{cont00} by  the
thermoelectrical force  $\pi^2 \nu {\nabla} T^2  /6mN = \pi^2 T {\nabla} T/3m\mu,$  whose  contribution
is   suppressed in the highly degenerate electron gas. Let us estimate this force in the lowest order in $T/\mu.$  To this end,  we neglect l.h.s. of Eq.~\eqref{T}  (which is small compared to its r.h.s. due to the same parameter $T/\mu$), thus arriving to  a balance equation between Joule heating and phonon cooling: $mv^2/\tau=(T-T_0)/\tau_{ph}.$
Hence,
the thermoelectrical force becomes $(\pi^2 T \tau_{ph}/3\mu\tau)\nabla v^2. $ Comparing this force with
the term $(\boldsymbol{v} \nabla)\boldsymbol{v},  $ we  conclude that the former
is negligible  provided that $\mu/T \gg \tau_{ph}/\tau.$  Assuming that this inequality is fulfilled,
  we are left with the system of  the  hydrodynamic equations for velocity and concentration:
 \BEA
&&
\label{eq:cont}
 \frac{{\partial n}}{{\partial t}} +  \frac{\p v_x}{\p x }=-\frac{\p (nv_x) }{\p x},
\\
&&\label{eq:hydro}
 \frac{\p v_x}{\p t} + \frac{v_x}{\tau } +s^2 \frac{\p n}{ \p x}=  a_x- v_x\frac{\p v_x}{\p x},
\\
&&\frac{{\partial {v_y}}}{{\partial t}}  + \frac{v_y}{\tau} = a_y - v_x \frac{\p {v_y}}{\p x},
\label{eq:hydro1}
\EEA
where $n=(N-N_0)/N_0$ and
$s=\sqrt{\frac{N_0}{m}\left(\frac{e^2}{C}+\frac{1}{\nu}\right)}
$
is the plasma wave velocity.

The r.h.s. of Eqs.~\eqref{eq:cont}, \eqref{eq:hydro}, and  \eqref{eq:hydro1} includes  perturbation  $\boldsymbol{a}$ as well as nonlinear  terms.  Assuming that $\boldsymbol{a}$ is small, one can search a solution as a perturbation series over  $\boldsymbol{a}:$
$n=n^{(0,1)}+n^{(1,0)}+\ldots, \hspace{3mm}
\boldsymbol{v}=\boldsymbol{v}^{(0,1)}+\boldsymbol{v}^{(1,0)}+\ldots$
Here the two indices denote the order of smallness with regard to
$\boldsymbol{e}$ and
$V_0,$ respectively.   The nonzero {\it dc} current
$\boldsymbol{j} = -e N_0{\big\langle {(1+n)\boldsymbol{v}} \big \rangle _{t,x}},
$
  appears in the third order   with respect to $\boldsymbol{a}$
(second order  in $\boldsymbol{e}$ and first order in $V_0$): $\boldsymbol{j}\approx \boldsymbol{j}^{(2,1)}$ (here $\langle\ldots \rangle_{t,x}$ stands for time and space averaging \cite{comment1}).
Importantly, Eqs.~\eqref{eq:cont} and \eqref{eq:hydro}  can be solved   independently from the  decoupled    Eq.~\eqref{eq:hydro1} [the latter can be
solved after the solution of  Eqs.~\eqref{eq:cont} and \eqref{eq:hydro} is  found].
  The details of calculations are presented in the Supplementary material. Here we  estimate  one  of the terms contributing to the $j_x^{(2,1)}$  in order to clarify the key points of  derivation.

The  static potential \eqref{eq:V}  leads to density  modulation $n^{(0,1)} \propto V_0 \cos(qx)  .$ The homogeneous part of the field  \eqref{eq:E} does not affect concentration but  leads to the Drude peak in the velocity: $v_x^{(1,0)}\propto E_{0x}[ e^{i\omega t}(i\omega+1/\tau)^{-1} + h.c.]$ (we omit here inhomogeneous contribution).   Substituting these equations into  nonlinear term $\p \left[n^{(0,1)} v_x^{(1,0)} \right] / \p x$  in the  r.h.s. of Eq.~\eqref{eq:cont},  and solving Eqs.~\eqref{eq:cont} and \eqref{eq:hydro} we find that velocity in the order (1,1) exhibits   plasmonic resonances  as well as  the Drude peak:  $v_x^{(1,1)}\propto  E_{0x} V_0 \cos(qx)  \times \left[e^{i\omega t}(i\omega +1/\tau)^{-1}(\omega^2-\omega_q^2 - i\omega/\tau)^{-1} + h.c.  \right ].$ Here $\omega_q=s q$ is the plasma wave frequency. In turn, nonhomogeneous part of the field \eqref{eq:E} also excites the plasmonic resonances, thus leading to density correction $n^{(1,0)} \propto E_{0x} h_x \sin(qx+\varphi)\left[e^{i\omega t} (\omega^2-\omega_q^2 - i\omega/\tau)^{-1} + h.c.  \right].$  Combining these equations,  we find   that there exists non-vanishing   correction to the {\it dc} current in the order (2,1):
$$j_x^{(2,1)} \propto    \left \langle n^{(10)} v_x^{(1,1)} \right\rangle_{t,x}\propto \frac{\tau} {1+\omega^2\tau^2} \frac{   \sin\varphi}{(\omega^2-\omega_q^2)^2 + \omega^2/\tau^2} .$$
A more detailed calculations presented in Supplementary material yield:
\BEA \label{eq:current-x}
&& j_{x}= j_{0x}\frac{2\omega _q^5 \tau  }{(1 + \omega ^2 \tau^2)[(\omega ^2 - \omega _q^2)^2 +\omega ^2/\tau^2  ]  },
\\ \label{eq:current-y}
&&j_{y} = j_{0y} \frac { \omega_q^3[(\omega ^2 - \omega _q^2)\tau \cos \theta  + \omega \sin \theta] }{(\omega ^2 - \omega _q^2)^2 +\omega ^2/\tau^2 }.
\EEA
Here $j_{0x}=e^4V_0N_0E_{0x}^2h_x \sin \varphi /(4m^3s^3 \omega_q^2)$ and $j_{0y}=-e^4V_0N_0E_{0x}E_{0y}h_y \sin \varphi /(4m^3s^3 \omega_q^2)$ are frequency- and disorder-independent currents that are proportional to   asymmetry factor $\sin\varphi$ and are sensitive  to the  polarization of the radiation. We note that the finite value of $j_y$ implies that electric circuit is closed in $y$ direction. For disconnected circuit, the voltage would develop instead, which is analogous to the Hall voltage and thus can depend on geometry of the system.

\begin{figure}[ht!]
     \leavevmode \epsfxsize=5.5cm
 \centering{\epsfbox{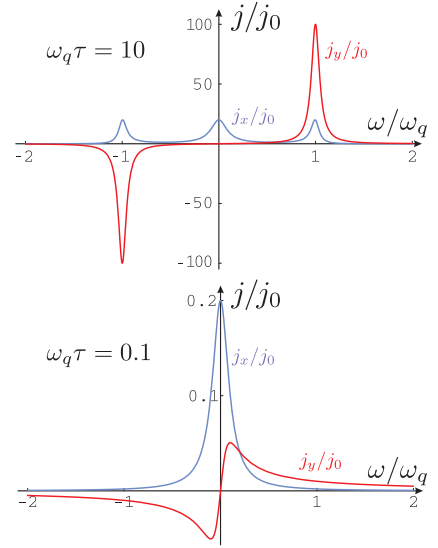}}
 \caption{Frequency dependence of current components in the   resonant (upper panel, $\omega_q\tau=10$) and non-resonant (lower panel, $\omega_q\tau=0.1$) cases for circular  polarization ($\theta=\pi/2,~E_{0x}=-E_{0y}, ~h_x=h_y, ~j_{0x}=j_{0y}=j_0. $)}
\vspace{-3mm}
\label{Fig1}
\end{figure}

As seen from Eqs.~\eqref{eq:current-x} and \eqref{eq:current-y}, there are two different regimes  depending on the plasmonic quality factor $\omega_q\tau.$
For $\omega_q\tau \gg 1,$  the response is peaked both  at $\omega=0$ and at  $\omega \simeq \omega_q$  within the frequency window $\sim 1/\tau.$
In the vicinity of the plasmonic resonance $\omega\simeq \omega_q,$ one can simplify Eqs.~\eqref{eq:current-x} and \eqref{eq:current-y}:
\BEA \label{jxfinal1}
&&j_{x} \approx j_{0x} \frac{2\omega_q\tau}{1+4(\omega-\omega_q)^2\tau^2}
 \\\label{jyfinal1}
&&j_{y}  \approx j_{0y} \frac{\omega_q^2\tau^2[\sin\theta + 2(\omega-\omega_q)\tau \cos\theta]}{1+4(\omega-\omega_q)^2\tau^2}.
\EEA

In the opposite non-resonant case, $\omega_q\tau \ll 1,  $ we find
$$j_x \approx \frac{2\omega_q \tau j_{0x} }{1+\omega^2 \tau_M^2 }, ~j_y \approx \frac{\omega_q \tau ( \omega \tau_M \sin \theta -\cos \theta) j_{0y}}{1+\omega^2 \tau_M^2 }
$$
where the width of the response,
$
{1}/{\tau_M}= \omega_q^2\tau,
$
is determined by the inverse time of the  charge spreading at the distance $ \sim q^{-1}$ (Maxwell relaxation time).

In the resonant regime  $\boldsymbol{j}$ is much larger than  in  the non-resonant case
(due to the largeness of  $\omega_q \tau$) and   shows sharp resonant dependence on $\omega$ (see Fig.~2).
Hence, excitation  of plasmons  leads to a dramatic  enhancement of the rachet effect.
Note that for $\theta=\pm \pi/2,$ $j_y$ changes its sign with
the sign of $\theta$, i.e. at switching between right and left circular polarizations. Thus, our results predict strong helicity effect - the circular polarization of the incident light determines the direction of $j_y.$ Remarkably, for clean systems  transverse component of the current, $j_y^{max}/j_{0y} \sim (\omega_q \tau) j_x^{max}/j_{0x},$  might be  much larger than the  longitudinal one  provided that $\omega_q\tau$ is sufficiently large. It worth  also stressing that for $\theta=\pm \pi/2$ transverse current remain finite in the dissipationless limit $\tau \to \infty:$ $j_y \to \pm j_{0y} \omega_q^3\omega/(\omega^2-\omega_q^2)^2.$

To conclude,  we predicted  a
  dramatic enhancement of the ratchet effect due to the  excitation of  plasmonic resonances.  We identified a helicity-dependent contribution to the ratchet current and  found that this contribution increases with decreasing the static disorder and saturates in the limit $\tau \to \infty.$
We also demonstrated    that the  non-resonant ratchet current  is  sharply peaked at zero frequency within the inverse Maxwell relaxation time.

We thank S. Ganichev, L. Golub, A. Muraviev, and V. Popov  for stimulating discussions. The
work has been supported
by
grant FP7-PEOPLE-2013-IRSES of the EU network Internom and by RFBR.

\begin{center} \section{ Supplementary material}
\end{center}

In this Supplementary material, we  present a rigorous derivation of {\it dc} current in the channel  
based on iteration of     Eqs.~\eqref{eq:cont}, \eqref{eq:hydro}, and  \eqref{eq:hydro1}
with respect to $\boldsymbol{a}$.
Non zero response  appears in the order (2,1) and can be written  as a sum of terms arising at different steps of iterations:
\BEA \label{j}
&&\boldsymbol{j}^{(2,1)}=-eN_0 \big \langle n^{(2,0)} \boldsymbol{v}^{(0,1)} + n^{(1,1)} \boldsymbol{v}^{(1,0)}
\\ &&  + n^{(1,0)} \boldsymbol{v}^{(1,1)} + n^{(0,1)} \boldsymbol{v}^{(2,0)} + \boldsymbol{v}^{(2,1)} \big \rangle_{t,x}. \nonumber
\EEA

Next, we calculate all terms entering Eq.~\eqref{j} separately for $j_x$  and $j_y.$
$$$$
\subsection{Calculation of $j_x.$}
We start  with calculation of  the $x-$component of the current.
In the orders (0,1) and (1,0)  the nonlinear terms in the r.h.s. of Eqs.~\eqref{eq:cont} and \eqref{eq:hydro} are absent, so that we are left with linear equations, whose solution  yields $n^{(0,1)}, n^{(1,0)},v_x^{(0,1)}, v_x^{(0,1)}.$  Next,
we substitute this solution into nonlinear terms and make the next iteration which yields the  terms  of the orders  (2,0), (1,1), and (0,2) (all terms in the second order in $\boldsymbol{a}$).      The solution in the order $(i,j)$  can be simply  written in the matrix form in the $(Q,\Omega)$ domain
\be \label{res}
\left(\hspace{-1.5mm}\begin{array}{c}n \\ v_x \end{array}\hspace{-1.5mm} \right)_{Q,\Omega}^{(i,j)}\hspace{-2mm}=\hspace{-0.5mm}\frac{i}{\Omega^2-\omega_Q^2+i\Omega/\tau}\hspace{-1mm}
\left[ \hspace{-1mm}                                                                 \begin{array}{cc}\Omega+i/\tau & Q \\s^2Q & \Omega \\  \end{array}  \hspace{-1mm}  \right]\hspace{-1mm}\left(\hspace{-1mm}\begin{array}{c}J_n \\ J_x \end{array}\hspace{-1mm} \right)_{Q,\Omega}^{(i,j)}.
\ee
Here  $\omega_Q=sQ$ is the plasma wave frequency and  $J_n^{(i,j)}$ and $J_x^{(i,j)}$ are
 the r.h.s.  of Eqs.~\eqref{eq:cont} and \eqref{eq:hydro}, respectively, in the order $ij.$
Due to nonlinearity of the problem,  frequency $\Omega$ and the wave vector $Q$ arising at each step of iterations are discrete and given by harmonics of the $\omega$ and $q,$ respectively: $\Omega=M \omega,~Q=K q,$ where $M$ and $K$ are  integer numbers. For $\Omega= \pm \omega, ~Q= \pm q,$ and $\omega \approx\omega_q$ there appear  a plasmonic  resonances in the ratchet response.

 In the order (0,1), we have: $J_n^{(0,1)}=0, ~J_{x}^{(0,1)}=-(eqV_0/m)\sin(qx).$ Using Eq.~\eqref{res} (with $\Omega=0$ and $Q=\pm q$),  we find
\BEA &&  v_x^{(0,1)} = 0,\label{v01}
\\ &&n^{(0,1)} = \frac{eV_0}{ms^2}\cos ( qx ). \label{n01}
\EEA

Next, we find  $J_n^{(1,0)}=0, ~J_x^{(1,0)}=-(eE_{0x}/m)\cos(\omega t)[1+h_x\cos(qx)],$ so that there are two types of terms: with $\Omega=\pm \omega, ~Q=0$ and with  $\Omega=\pm \omega, ~Q=\pm q.$ From Eq.~\eqref{res} we get
\BEA
&& n^{(1,0)}=\frac{eE_{0x}h_x q\sin(qx+\varphi) e^{i\omega t}}{2m (\omega^2-\omega_q^2 -i\omega/\tau)}+h.c.,\label{n10}
\\
&&v_x^{(1,0)}=-\frac{eE_{0x}h_x \omega\cos(qx+\varphi)e^{i\omega t}}{2im(\omega^2-\omega_q^2-i\omega/\tau) } \label{v10}
\\\nonumber
 &&-\frac{eE_{0x}e^{i\omega t}}{2 m (i\omega +1/\tau)} +~h.c..
\EEA

Next, we substitute obtained solutions in the nonlinear terms and find
\BEA
&&J^{(1,1)}_x=-\frac{\p \big[v_x^{(1,0)}v_x^{(0,1)}\big]}{\p x}=0, \label{Jv11}
\\
&&J_n^{(1,1)}= - \frac{\p\left[n^{(1,0)}v^{(0,1)}+n^{(0,1)}v^{(1,0)}\right]}{\p x}                     \label{Jn11}
\\ \nonumber
&&=\frac{e^2E_{0x} V_0 q\sin(qx)e^{i\omega t}}{2m^2s^2(i\omega+1/\tau)} + h.c.+\ldots.
\EEA
 Here and in what follows $(\ldots)$ stands for terms oscillating in space with the wave vector $2q.$ We skip them since    their contribution drops out from $j_x^{(2,1)}$ after space averaging.
Substituting Eqs.~\eqref{Jv11} and \eqref{Jn11} in Eq.~\eqref{res}, we find
\BEA
&&n^{(1,1)}=-\frac{e^2E_{0x} V_0 q\sin(qx)e^{i\omega t}}{2m^2s^2(\omega^2-\omega_q^2-i\omega/\tau)}+h.c.+\ldots,  \label{n11}
\\
&& v_x^{(1,1)}=\frac{e^2E_{0x} V_0 q^2\cos(qx)e^{i\omega t}}{2 m^2(i\omega+1/\tau)(\omega^2-\omega_q^2-i\omega/\tau)}   \label{v11}
\\ \nonumber && + h.c. + \ldots.
\EEA
As seen from Eq.~\eqref{j} what is left to be done is the calculation of $v_x^{(2,1)}$ and $v_x^{(2,0)}.$

Let us first demonstrate that
  \be
  \label{v21}
  \left \langle v_x^{(2,1)} \right \rangle_{t,x}=0.
  \ee
  To this end, we notice that all terms  entering  Eq.~\eqref{eq:hydro}  except $v_x/\tau$   can be written as  derivatives over  $t$ or $x.$  Hence, averaging this equation over time and distance we find that
  $\left \langle v_x \right \rangle_{t,x}=0,$ and, consequently,  $\left \langle v_x^{(i,j)} \right \rangle_{t,x}=0$ for any $i$ and $j.$

In order to find $\left \langle v_x^{(2,0)}n^{(0,1)} \right \rangle_{t,x},$ we first write Eq.~\eqref{eq:cont}  in the order (2,0),
\be
\frac{\p n^{(2,0)}}{\p t} + \frac{\p v_x^{(2,0)}}{\p x}=-\frac{\p \left[n^{(1,0)}v_x^{(0,1)}+n^{(0,1)}v_x^{(1,0)}\right]}{\p x}.
\ee
   Then, we average this equation over time.  Since $v_x^{(0,1)}=0,$  we get
  \be
  \frac{\p\left[ v^{(2,0)}_x+ n^{(1,0)}v_x^{(1,0)}\right]}{\p x}=0.
  \ee
   Hence
 \be
 \left\langle  v_x^{(2,0)} \right \rangle_t=-  \left\langle  n^{(1,0)}v_x^{(1,0)} \right \rangle_t + \xi(t),
 \ee
where $\xi(t)$ is function of $t$ only.  Having in mind that $n^{(0,1)}$ depends on $x$ only and  $\left \langle n^{(0,1)}\right \rangle_x=0, $ we  get
\be \left \langle v_x^{(2,0)}n^{(0,1)} \right \rangle_{t,x}= - \left \langle n^{(0,1)} n^{(1,0)}v_x^{(1,0)}\right \rangle_{t,x}. \label{v20}\ee

Now, we can calculate  $j_x.$ Using  Eqs.~\eqref{v10} and \eqref{n11} we find  $\left \langle v_x^{(1,0)}n^{(1,1)} \right \rangle_{t,x}=0.$ Since $v_x^{(1,0)}=0,$ we conclude that only two terms contribute to $j_x:$
\be
j_x^{(2,1)}=-eN_0 \left \langle  n^{(1,0)}v^{(1,1)}_x  -n^{(0,1)}n^{(1,0)}v^{(1,0)}_x \right \rangle_{t,x}.
\label{jx0}
\ee
Substituting  Eqs.~\eqref{n01}, \eqref{n10}, \eqref{v10}, and \eqref{v11}  into Eq.~\eqref{jx0} and averaging over $t$ and $x,$ we find  that both terms in Eq.~\eqref{jx0} yield equal contributions. The total current $j_x\approx j^{(2,1)}_x$ is given by Eq.~\eqref{eq:current-x}  of the main text.

\subsection{Calculation of $j_y$}

In this subsection, we find transversal component of current  solving Eq.~\eqref{eq:hydro1} by iterations with respect to small $\boldsymbol{a}.$  In the order $(i,j)$ solution  in $(Q,\Omega)$ space reads
\be  \label{vyQW}
v_y^{(i,j)}(Q,\Omega)=\frac{J_y^{(i,j)}(Q,\Omega)}{1/\tau-i\Omega},
\ee
where $J_y^{(i,j)}(x,t)$ is the r.h.s. of Eq.~\eqref{eq:hydro1}.

In the first order with respect to $\boldsymbol{a}$ we find
\BEA
&& J_y^{(0,1)}(x,t)=0,\\
&&J_y^{(1,0)}(x,t)=a_y= -\frac{eE_{0y}}{m} \cos(\omega t +\theta)
\\\nonumber &&\times [1 + h_y\cos(qx+\varphi)] .
\EEA
Using Eq.~\eqref{vyQW}  we get
\BEA
&& v^{(0,1)}_y=0, \label{vy01}
\\
&&v_y^{(1,0)}=\frac{eE_{0y}[1+h_y\cos(qx+\varphi)]e^{i(\omega t +\theta)}}{2m(i\omega+1/\tau)}  \label{vy10}\\ \nonumber
&& +~h.c.
\EEA

Next, we consider terms of the second order  with respect to $\boldsymbol{a}$.
In the order (1,1) we get
\be
J_y^{(1,1)}=-v_x^{(1,0)} \frac{\p v_y^{(0,1)}}{\p x} -v_x^{(0,1)} \frac{\p v_y^{(1,0)}}{\p x}=0,
\ee
  so that
 \be
 v_y^{(1,1)}=0 \label{vy11}
 \ee
Having in mind Eqs.~\eqref{vy01} and \eqref{vy11} we find  from Eq.~\eqref{j}
\be {j}_y^{(2,1)}=-eN_0 \big \langle  n^{(1,1)} v^{(1,0)}_y
  + n^{(0,1)} {v}_y^{(2,0)} + {v}_y^{(2,1)} \big \rangle_{t,x}.
  \label{jy}
  \ee
  From Eqs.~\eqref{n11} and \eqref{vy10} we obtain
  \BEA \label{n11vy10}
  &&\big \langle  n^{(1,1)} v^{(1,0)}_y \big \rangle_{t,x}= -\frac{e^3E_{0x}E_{0y} h_y V_0 q\sin \varphi}{4m^3s^2 (\omega^2+1/\tau^2)}\\ \nonumber
  && \times\frac{[(\omega_q^2/\tau) \cos\theta +\omega (\omega_q^2-\omega^2-1/\tau^2)\sin\theta ]}{(\omega^2-\omega_q^2)^2+\omega^2/\tau^2}.
  \EEA
  As follows from Eq.~\eqref{n01}, $n^{(0,1)}$ does not depend on $t.$ Thus, while calculating  contribution of the second term in Eq.~\eqref{jy} we can first average in time ${v}_y^{(2,0)}.$ Averaging over time  Eq.~\eqref{eq:hydro1}, we find in the order (2,0)
  \be
  \big\langle v_y^{(2,0)} \big\rangle_t= \tau\left\langle v_x^{(1,0)} \frac{\p v_y^{(1,0)}}{\p x}\right\rangle_t.
  \ee
Hence,
  \BEA
  && \label{vy20n01}
  \left\langle n^{(0,1)}{v}_y^{(2,0)}\right\rangle_{t,x}= \tau\left\langle n^{(0,1)} v_x^{(1,0)} \frac{\p v_y^{(1,0)}}{\p x}\right\rangle_{t,x} \\ \nonumber
  && =\frac{e^3 E_{0x}E_{0y} h_y V_0 q \tau \sin \varphi \cos \theta }{4 m^3 s^2(\omega^2+1/\tau^2)}.
  \EEA
Here we used Eqs.~\eqref{n01}, \eqref{v10}, and  \eqref{vy10}.

Analogously,  averaging over time and distance Eq.~\eqref{eq:hydro1} in the order (2,1), and using Eqs.~\eqref{v11} and \eqref{vy10}, we get
\BEA \label{vy21}
&&\left\langle v_y^{(2,1)} \right\rangle_{t,x}= \tau\left\langle v_x^{(1,1)} \frac{\p v_y^{(1,0)}}{\p x}\right\rangle_{t,x}
\\ \nonumber
&&=-\frac{e^3 E_{0x}E_{0y} h_y V_0 q^3 \tau \sin \varphi }{4 m^3 (\omega^2+1/\tau^2)}
\\ \nonumber
&&\times \frac{[(\omega^2-\omega_q^2) \cos\theta +(\omega/\tau)\sin\theta ]}{(\omega^2-\omega_q^2)^2+\omega^2/\tau^2}
.
\EEA

Finally, substituting Eqs.~\eqref{n11vy10}, \eqref{vy20n01},  and \eqref{vy21} into Eq.~\eqref{jy},
we find that $j_y \approx j_y^{(2,1)}$ is given by Eq.~\eqref{eq:current-y} of the main text.


\begin{thebibliography}{8}
\bibitem{comment0}   Note that  for  realistic systems the density and field
perturabtions  can not be described  by simple harmonic fucnctions and  involve infinite number of harmonics (see Ref.~\cite{ivchenko}). Also, the  field amplitude $\boldsymbol{e}$ is much smaller than the  amplitude of the external field due to the screening by the gate electrodes.  However, the simplified  model based on Eq.~\eqref{eq:V} and \eqref{eq:E} with phenomenological parameters $V_0,$ $E_{0x},$ and $E_{0y}$    captures the key physics of the problem and is sufficient for clarifying the basic concept of the plasmon-enhanced ratchet.
\bibitem{comment1} In fact,  averaged in time $\boldsymbol{j}$ does not depend on $x,$ so that it is sufficient to make  averaging over $t$ only. However, calculations are strongly simplified if we  make also  $x$-averaging in all contributing  terms.
\end{thebibliography}
\end{document}